\documentstyle[preprint,revtex]{aps}
\def\o{{\it odd }}
\def\e{{\it even }}

\def\bk{{\bf k}}
\def\bp{{\bf p}}

\def\bS{{\bf S}}
\def\bK{{\bf K}}

\def\pmb#1{\setbox0=\hbox{#1}
     \copy0\kern-\wd0 \kern-0.01em\copy0\kern-\wd0
\kern-0.01em\copy0\kern-\wd0 \kern0.03em\copy0\kern-\wd0
\kern0.01em\copy0\kern-\wd0 \kern-0.04em\raise0.018em\copy0\kern-\wd0
\kern0.01em\raise0.018em\copy0\kern-\wd0
\kern0.01em\raise0.018em\copy0\kern-\wd0
\kern0.01em\raise0.018em\box0}
\def\bbf#1{\mathchoice{\pmb{$\displaystyle #1$}}{\pmb{$\textstyle #1$}}%
     {\pmb{$\scriptstyle #1$}}{\pmb{$\scriptscriptstyle #1$}}}

\begin{document}

\draft
\preprint{LA-UR-93-2992}

\begin{title}
Properties of Odd Gap Superconductors \footnotemark \footnotetext{ Invited talk
at the  Conference on Strongly Correlated Electron Systems, San Diego, 1993}
\end{title}

\author{A. V. Balatsky\footnotemark} \footnotetext{ Email: avb@viking.lanl.gov;
tel: (505)665.13.14; fax: (505)665.29.92}
\begin{instit}
 Theoretical Division\\ Los
Alamos National Laboratory, Los Alamos, NM 87545\footnotemark \footnotetext{
Also  at the Landau
Institute for Theoretical Physics, Moscow, Russia}
\end{instit}
\author{Elihu Abrahams}
\begin{instit}
Serin Physics Laboratory, Rutgers University, P.O. Box
849, Piscataway, NJ 08855
\end{instit}
\author{D.J. Scalapino}
\begin{instit}
Department of Physics, University of California, Santa Barbara, CA  93106
\end{instit}
\author {J.R. Schrieffer}
\begin{instit}
Department of Physics, Florida State University, Tallahassee, FL 32310
\end{instit}

\receipt{August 12, 1993}
\begin{abstract}
A new class of superconductors with the gap function \o under time reversal is
considered. Some of the physical properties of these superconductors such as
the Meissner effect, composite condensate,  gapless spectrum and transition
from the \o gap superconductor to the BCS state at lower temperatures are
discussed.

\end{abstract}


The discovery of high temperature superconductivity has stimulated  new
approaches to the  phenomenon of superconductivity in the strongly correlated
systems.
Recently a new class of singlet superconductors has been proposed in  \cite{BA}
with a remarkable feature that the gap function $\Delta(k, \omega)$ is {\em
odd} under inversion of momentum and frequency separately. The original
motivation for considering this new class was the fact that this pairing
channel is insensitive to the short range Coulomb repulsion and thus might be a
relevant pairing channel in strongly correlated systems \cite{Allen}. This new
class of superconductors was called {\em odd frequency} gap superconductors,
hereafter {\em odd} gap superconductors.  In the context of the superfluidity
of $^3He$ the analogous class of spin triplet \o gap superconductors was
considered by Berezinskii some time ago \cite{Ber}.

Here we will discuss some  general properties of the \o gap superconductor,
such as the  Meissner effect, composite operator description of \o gap
superconductors and the second phase transition from the \o gap superconductor
to the BCS superconductor \cite{G4}.



We begin with the symmetry equation, from which the existence of the \o class
of superconductors follows \cite{BA,Ber}. Consider spin singlet anomalous
function $F({\bf k}, \omega_n)~=~{1\over 2}\sum_{\alpha,\beta}\int d{\bf
r}~\int^
{\beta}_{-
\beta}~d\tau~e^{i\omega_n\tau}~e^{-i{\bf k}\cdot{\bf r}}~\langle
T_\tau\,c_\alpha ({\bf r},\tau)~c_\beta (0,0)\rangle
(i\sigma^{y})_{\alpha
\beta}$ with $(i\sigma_{y})_{\alpha
\beta}$
being  a spin metric tensor, $\tau$ being  the Matsubara time and $\beta =
1/T$.  Note that the anomalous Green's function is explicitly written
in a general spin-singlet form. The same set of definitions holds for a gap
function $\Delta(\bk, \omega)$. The {\em only} constraint on the possible
symmetry of $F(\bk, \omega)$ and
$\Delta(\bk, \omega)$ comes from the fact that the $c_\alpha ({\bf r},\tau)$
are fermions, and using the definition of the time ordering operator one can
show that \cite{BA}:
\begin{equation}
F({\bf k},\omega_{n})~=~F(-{\bf k}, -\omega_{n}),~~\Delta ({\bf
k},\omega_n)~=~\Delta (-{\bf k}, -\omega_n)
\end{equation}
Apart from the standard BCS (\e gap) solution of this constraint in terms of
the $P$-even (parity) and $T$ even (time reversal), we immediately find that
there is a new class of solutions with $P$-odd and $T$-odd behavior of the
anomalous function $F$ and gap $\Delta$ \cite{BA}. In this new class the
singlet $p$ wave pairing state is allowed, inverting relation between spin
state and orbital parity. There is no contradiction with the Pauli principle
since the equal time wave function vanish in this state.

It is  clear that the physical properties of the \o gap superconductors are
different from we are familiar with from  BCS superconductors. The main
difference comes from the fact  that the gap function $\Delta(\bk, \omega)$, as
well as anomalous function $F(\bk, \omega)$, are \o functions of frequency
(time). We    list  some of the peculiar properties of \o gap superconductors
\cite{BA,G4}:

1){\em Broken P and T symmetries}.  By definition singlet  \o gap
superconductor breaks $T$ and $P$.

2){\em Quasiparticle spectrum}. The quasiparticle  spectrum of the \o gap
superconductor is {\em gapless} because the  equal time gap vanish $\Delta(\bk)
= \sum_n \Delta ({\bf k},\omega_n) = 0$. The pole of the Green's function
defines the quasiparticle spectrum on the superconducting state at   $\omega^2
- \epsilon_k^2 -\Delta^2(\bk, \omega) = 0$. For linearized $\Delta(\bk, \omega)
= a_{\bk } \omega$ at $\omega \rightarrow 0$ we find that the quasiparticle
spectrum is given by the renormalized free particle dispersion \begin{equation}
\omega_k = \epsilon_k (1 - a_{\bk}^2)^{-1/2}
\end{equation}
so that quasiparticle spectrum remains gapless in the \o gap superconductor.
This also implies that the low temperature specific heat and the penetration
depth in this superconductor will be  power law functions  of temperature.

3){\em Order parameter and composite operator}. The \o gap superconductor has a
new  order parameter. For any \o in time gap function for both spin channels,
assuming analyticity of $\Delta_{\alpha\beta}$ in small relative pair field
time $\tau$, we have $ \Delta_{\alpha\beta}^{\it odd} = 2\tau
\Delta_{\alpha\beta}^{(1)} +
{\cal O}(\tau ^3)$. Thus the {\em slope} $\Delta_{\alpha\beta}^{(1)} =  \langle
 \dot c_{\alpha,{\bf k}} c_{\beta,-{\bf k}}- c_{\alpha,{\bf k}} \dot
c_{\beta,-{\bf k}}\rangle
$ of the gap function at $\tau = 0$ can be taken as the {\em order parameter}
for \o gap superconductor.  To calculate the time derivative of any operator we
need to use the Hamiltonian. Let us  consider a Kondo lattice  model with
conduction band in which the superconducting instability takes place and with
the set of localized spins in the second band: $ H = \sum_{\bk, \sigma}
\epsilon_{\bk} c^{\dag}_\sigma(\bk) c_\sigma(\bk) + J \sum_{\bk, \bp}
c^{\dag}_\alpha(\bk - \bp) c_\beta(\bk) \bbf{\sigma}_{\alpha \beta}\bS_{\bp} +
H_{spin}$, where $H_{spin}$ is the  Hamiltonian, which descibes spins  with the
 propagator $D({\bf r}, \tau) = -\langle T_{\tau} \bS ({\bf r}, \tau) \bS (0,0)
\rangle$.  Using equation of motion for $\dot c_{\alpha}(\bk)$  one finds for
the  Berezinskii \o gap triplet state with ${\bf K} = (\sigma^y
\bbf{\sigma})_{\alpha \beta} \Delta_{\alpha\beta}^{(1)}$:
\begin{equation}
{\bf K}  =   J \sum_{\bk, \bp}\langle c_\alpha(\bk - \bp)
c_\beta(\bk)\sigma^y_{\alpha \beta} \otimes \bS_{\bp} \rangle
\label{op}
\end{equation}
The order parameter in Eq.(\ref{op}) is a {\em composite operator} describing
the pairing in the \o gap superconductor.  The spin triplet condensate occurs
in the form of a bound state of the  $S = 0$ Cooper pair (the first part in the
product in Eq.(\ref{op})) and the $S = 1$ spin boson $\bS$. Analogous
considerations  for the spin singlet shows that {\em composite operator}
describes the  spin triplet Cooper pair bound with spin $S = 1$ boson to yield
a total singlet. This  form of the {\em composite operator} was proposed
originally by Berezinskii \cite{Ber} and later considered in
\cite{EK,BB,G4,Coleman}.

 Thus we are led to consider the susceptibility in the {\em composite operator}
channel to reveal the \o gap instability in this Hamiltonian formalism. The
simplest minimal model which leads to the mean field instability with the {\em
composite } condensate is  quadratic in  $\bK$ and  is analogous to the BCS
Hamiltonian for the weak coupling theory:
\begin{equation}
H =  \sum_{\bk, \sigma} \epsilon_{\bk} c^{\dag}_\sigma(\bk) c_\sigma(\bk) + V
\sum_{\bk, \bp} c^{\dag}_\alpha(\bk - \bp) c_\beta(\bk) \sigma^y_{\alpha
\beta}\bS_{\bp}\cdot \bK  + |\bK|^2/2
\end{equation}
The order parameter $\bK$ describes the new {\em anomalous vertex} in the
superconducting state with two fermions and one boson vanishing in the
condensate, similar to the two fermion anomalous Green's function  in the  BCS
superconductor.

The diagram corresponding to the $\bK$ {\em anomalous vertex} is  shown in the
Fig.1. The selfconsistency equation for $\bK$ involves the {\em three particle}
 (two fermions and boson) scattering vertex $V$ and gives $\bK = T^2 \sum_{n,
m, \bp, \bk} G(\omega_n, \bp) G(\omega_m, \bk) D(\bp +\bk, \omega_n + \omega_m)
\bK$. Taking $D(\bp,\omega) \sim \delta(\omega)\delta(\bp)$ we recover the  BCS
selfconsistency equation, what corresponds to the case when spins are
ferromagnetically ordered and can be factorized in the Eq.(3).

The results we obtain from this phenomenlogical Hamiltonian are similar to the
predictions made  in  the Elaishberg theory for \o gap superconductor with
spin boson interaction   mediating   pairing in the \o frequency channel.

4){\em Critical coupling}. We find a critical coupling $g_{crit}  = N_0
V_{crit}  = O(1)$ for the \o frequency superconducting instability. This occurs
because the gap function is \o in time (frequency) and the phase space
available  for pairing interaction is greatly diminished at low frequency. The
BCS like infrared instability is directly related to the fact that the gap
function $\Delta(\bk, \omega)$ is nonzero at $\omega \rightarrow 0$. Thus for
any $\Delta(\bk, \omega) \rightarrow 0$, as $ \omega \rightarrow 0$  the
transition takes place only above some threshold \cite{G4}. The phase
transition into the \o gap superconducting phase is of the {\em second order}
with the order parameter $K$  having  the mean field exponent $\nu  = 1/2$:
$|K| \sim (T_c - T)^{1\over2}, |K| \sim (g - g_{crit})^{1\over2}$.

5){\em Meissner effect}. There is the   Meissner effect  for the \o gap
superconductors. We have at the  moment two distinct approaches to the
calculation of the Meissner effect in these superconductors. The straighforward
use of the normal and anomalous Green's functions in the current-current
correlator yields a negative Meissner effect for the spatially homogeneous
ground state in the vicinity of $T_c$ (see however \cite{Coleman}). On the
other hand using the {\em composite operator} description of the \o gap
superconductor, as in Eq.(4) it is easy to show  the existence of the positive
Meissner effect in the  homogeneous phase with momentum independent
$\bK$\cite{G4}. This difference reflects the choice of the spatial structure of
the ground state in {\em composite operator }.

6) {\em Transition from the \o gap superconductor to the BCS superconductor}.
The effective theory of \o gap superconductor with composite order parameter
has a  {\em second} phase transition into the BCS superconductor at lower
temperatures. The anomalous vertex $\bK$ is a charge $2e$ operator. Thus any
other condensate with charge $2e$ is allowed in this phase  and the transition
could take place  at any temperature below the first  transition into the \o
gap phase. To see this consider the {\em even} frequency spin singlet anomalous
Green's function $F_{BCS}({\bf r}, t) = - \langle T_\tau\,c_\alpha ({\bf
r},\tau)~c_\beta (0,0)\rangle \sigma^y_{\alpha \beta}$. The second order
expansion in  $\bK$ in the trace inside the brackets in $F_{BCS}$ yields the
term which is a source for the spin singlet BCS anomalous function (see Fig.2)
with:
\begin{equation}
F_{BCS}(\bk,n) =  -T \bK^2 G(\bk,n)G(-\bk,-n) \sum _{m, \bp}F_{BCS}^*(\bp,m)
D(\bp -\bk,m-n)
\end{equation}
It follows  that the spin triplet \o gap superconductor might be unstable at
lower tempeartures towards the {\em even} frequency spin singlet
superconductor. How relevant this new pairing channel is will  be investigated.
 On the other hand the direct exchange between particles  due to $D(\bk,
\omega_n)$ can produce only spin triplet \e gap and one has to compare the free
energies of different states. Because the symmetry of the BCS order parameter
is different from the symmetry of the \o gap, we generally expect the second
transition to occur at temperatures away from  the $T_c$ for the \o gap
superconductor.  The effective coupling in the BCS channel is proportional to
$\bK^2$ and is small near $T_c$ for the \o gap superconductor thus making
second transition temperature $T^{BCS}_c < T_c$.  The crucial difference of
Eq.(5) from the BCS selfconsistency equation is that the vertex is {\em
anomalous} and as a result we have $F_{BCS}$ on the l.h.s. related to
$F_{BCS}^*$ on the r.h.s. of Eq.(5), whereas in the BCS theory anomalous
function is expressed through itself in the selfconsistency equation. This also
means that the phase of $\bK$ is related to the the phase of $F_{BCS}$.

It is  possible that this hierarchy of transitions is an  inherent property of
any \o gap superconductor. It might be   that this phenomenon is related to the
second phase transition  observed in some  of the heavy fermions far inside
the superconducting phase.

\underline{Acknowledgments} A.V.B. is  grateful to P. Coleman, V. Emery and S.
Kivelson  for useful discussions.  This work was supported by the J.  R.
Oppenheimer fellowship, by the Department of Energy and by NSF-DMR.


\figure{The anomalous {\em vertex} corresponding to the {\em composite}
operator which describes the binding of the Cooper pair with boson in the \o
gap superconductor condensate.}

\figure{The second order in $\bK$ diagram which leads to the nontrivial
solution for BCS spin singlet gap and  to Eq.(5) for weakly momentum dependent
$\bK$. The vertex in this diagram is anomalous and does not conserve the number
of particles, which makes this diagram different from the gap equation in the
standard Eliashberg theory.}

\end{document}